\documentclass[aps,prb,twocolumn,superscriptaddress]{revtex4}

\usepackage{graphicx}	
\usepackage{amssymb}	
\usepackage{amsfonts}	
\usepackage{amsmath}	
\usepackage{xcolor}		
\usepackage[T1]{fontenc}	
\usepackage[utf8]{inputenc} 
\usepackage{makecell} 
\usepackage{booktabs} 

\usepackage[english]{babel} 	
\usepackage[colorlinks=true,linkcolor=blue,citecolor=blue]{hyperref} 

\usepackage{braket}		
\usepackage[load-configurations = abbreviations]{siunitx}	

\DeclareMathOperator{\Tr}{Tr}

\begin{document}

\title{Large enhancement of the thermoelectric power factor in disordered materials through resonant scattering}

\author{S. Th\'ebaud}
\email[E-mail: ]{simon.thebaud@univ-lyon1.fr}
\affiliation{Univ Lyon, Universit\'e Claude Bernard Lyon 1, CNRS, Institut Lumi\`ere Mati\`ere, F-69622, LYON, France}
\author{Ch. Adessi}
\affiliation{Univ Lyon, Universit\'e Claude Bernard Lyon 1, CNRS, Institut Lumi\`ere Mati\`ere, F-69622, LYON, France}
\author{G. Bouzerar}
\affiliation{Univ Lyon, Universit\'e Claude Bernard Lyon 1, CNRS, Institut Lumi\`ere Mati\`ere, F-69622, LYON, France}

\begin{abstract}
In the search for more efficient thermoelectric materials, scientists have placed high hopes in the possibility of enhancing the power factor using resonant states. In this study, we investigate theoretically the effects of randomly distributed resonant impurities on the power factor. Using the Chebyshev Polynomial Green's Function method, we compute the electron transport properties for very large systems (\SI{\sim e7}{atoms}) with an exact treatment of disorder. The introduction of resonant defects can lead to a large enhancement of the power factor together with a sign inversion in the Seebeck coefficient. This boost depends crucially on the position of the resonant peak, and on the interplay between elastic impurity scattering and inelastic processes. Strong electron-phonon or electron-electron scattering are found detrimental. Finally, the robustness of our results is examined in the case of anisotropic orbitals and two-dimensional confinement. Our findings are promising for the prospect of thermoelectric power generation.
\end{abstract}

\maketitle

\section{Resonant states in thermoelectric materials}

Over the past decades, the increasingly pressing need for clean energy sources and the realization that a huge proportion of the energy used worldwide is wasted in heat \cite{Forman2016} have prompted great interest in the perspective of developping efficient thermoelectric generation modules.\cite{Zheng2011,HamidElsheikh2014,Zhang2015,Zhu} The currently available devices, mainly based on (Bi,Sb)$_2$Te$_3$, PbTe or Si-Ge alloys, are not efficient enough to be used industrially on a large scale. \cite{Mahan2016} The efficiency of a thermoelectric module is limited by the temperature-averaged figure of merit $zT$ of both the n and p legs of the module, \cite{Zlatic:1667229,Goldsmid,Snyder2003} with
\[
zT = \frac{\sigma S^2}{\kappa} T, 
\]
in which $\sigma$ is the electrical conductivity, $S$ is the Seebeck coefficient and $\kappa$ is the thermal conductivity, often dominated by phonons in doped semiconductors. The use of thermoelectric devices on a large scale would require a figure of merit of at least 2 to 3, depending on the application area.\cite{Bell2008,Thekdi2015} The thermal averaging must be done between the temperatures of the heat source and that of the heat sink, meaning that the materials composing the legs should have a good figure of merit on a wide range of temperatures. Additionally, the efficiency and reliability of the device is partly determined by the compatibility between the two legs, so ideally they should be made of similar compounds, such as a single semiconductor host with a different doping.\cite{doi:10.1002/admt.201700256,Bell2008}

So far, most progress in boosting the thermoelectric figure of merit has been accomplished by lowering the thermal conductivity. \cite{Kanatzidis2010,Snyder2008} Enhancing the power factor (PF) $\sigma S^2$ is a much more formidable challenge due to the interplay between the electrical conductivity and the Seebeck coefficient. To go beyond the rigid band doping optimum, various band engineering strategies have been proposed, such as band convergence \cite{Pei2012,Yang2015a,Pei2011} and dimensional confinement of the electron gas.\cite{Hicks1993a,Heremans2005,Dresselhaus2007,Pichanusakorn2010} Another promising method to boost the PF consists in using resonant states. \cite{Heremans2012,Mahan7436,PhysRevLett.107.226601} The concept is illustrated in Fig.~\ref{fig1}. Dopants introduce impurity states inside the conduction band that hybridize with the extended states. This way, a sharp peak in the density of states (DOS) of the host compound is created, which alters the electronic transport properties when the Fermi level lies in its vicinity. An enhancement of the thermoelectric properties through resonant impurity states has been claimed in various compounds, such as Tl-doped PbTe, \cite{Heremans2008} Sn-doped Bi$_2$Te$_3$, \cite{Jaworski2009} In-doped SnTe, \cite{Zhang2013} Al-doped PbSe, \cite{Zhang2012} or Sn-doped $\beta$-As$_2$Te$_3$. \cite{Wiendlocha2018} The case of Tl-doped PbTe is controversial since first-principles calculations reproduced the experimental values for the Seebeck coefficient with a simple rigid-band shift of the pristine material. \cite{Singh2010} Subsequently, numerical studies using the Coherent Potential Approximation (CPA) coupled with first-principles calculations \cite{Minar2011} have been conducted to investigate the effects of random Tl doping in PbTe. \cite{WIENDLOCHA201633,PhysRevB.88.205205,PhysRevB.97.205203} However, transport properties were calculated in the absence of electron-phonon scattering, and the treatment of disorder by CPA methods is known to ignore the vertex corrections and thus localization effects. \cite{Economou} More importantly, no clear improvement in the PF over a rigid-band shift of the Fermi level was shown for Tl-doped PbTe. As it stands, there is still no consensus in the literature whether actual resonant enhancement of the thermoelectric properties have been observed experimentally.

In this study, we aim to clarify the general conditions required for a boost of the PF using resonant substitution impurities, all with a full treatment of the disorder and resonant scattering. In particular, we will investigate the influence of the impurity concentration $x$, the effects of inelastic scattering, and finally we will examine the case of anisotropic orbitals. 

\section{Disordered model Hamiltonian and methodology}

\begin{figure}
\includegraphics[width=0.9\columnwidth]{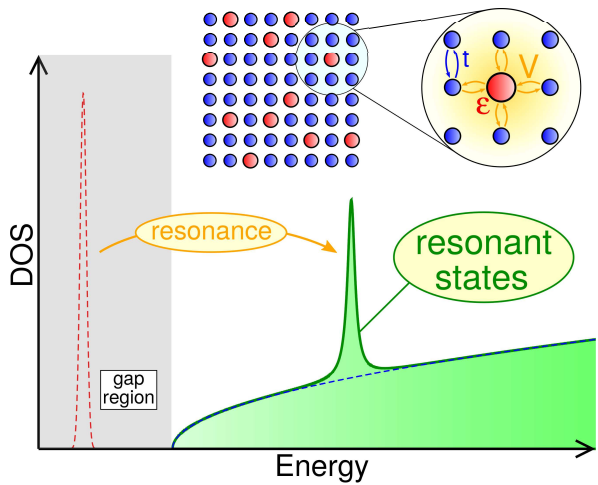}
\caption{Sketch of the density of states as a function of energy corresponding to a conduction band in the presence of resonant states. Inset: illustration of the Hamiltonian terms.}
\label{fig1}
\end{figure}

To keep the conclusions as general as possible, we consider a single-orbital tight-binding Hamiltonian featuring hopping terms $t$ between nearest neighbors on a cubic lattice. Here the charge carriers are electrons (n-type), but because of electron-hole symmetry, our results are valid for p-type materials as well. The resonant impurities are modelled by an on-site potential $\epsilon$ on the defect sites and an hybridization $V$ between the host and impurity sites. The position of the defects are chosen randomly. Hence, the Hamiltonian, as illustrated in the inset of Fig.~\ref{fig1}, reads:
\begin{align}
\label{Hamiltonian}
\hat{H} = & -t \sum_{\bf{\langle i,j \rangle},\sigma} \left( c^{\dagger}_{\bf{i},\sigma} c^{}_{\bf{j},\sigma} + c^{\dagger}_{\bf{j},\sigma} c^{}_{\bf{i},\sigma} \right) \\ 
				& + \epsilon \sum_{\bf{m},\sigma}  c^{\dagger}_{\bf{m},\sigma} c^{}_{\bf{m},\sigma} \; - V \sum_{\bf{\langle i,m \rangle},\sigma}  \left( c^{\dagger}_{\bf{i},\sigma} c^{}_{\bf{m},\sigma} + c^{\dagger}_{\bf{m},\sigma} c^{}_{\bf{i},\sigma} \right), \nonumber 
\end{align}
$\bf{i}$ runs over host sites, $\bf{m}$ over impurity sites, $\sigma$ is the electron spin and the brackets denote nearest neighbors only. Since the transport properties do not depend explicitely on $t$, we express the other parameters in units of $t$ (see below).

The electrical conductivity and the Seebeck coefficient at electron density $n$ and temperature $T$ write \cite{Ashcroft,PhysRevB.68.125210}
\begin{equation}
\sigma(n,T) = \int dE \left( -\frac{\partial f}{\partial E}\right) \Sigma(E),
\label{conductivite}
\end{equation}
and
\begin{equation}
S(n,T) = -\frac{1}{eT\sigma} \int dE \left( -\frac{\partial f}{\partial E}\right) (E-\mu) \, \Sigma(E),
\label{Seebeck}
\end{equation}
with $f(E,\mu,T)$ the Fermi distribution. The chemical potental $\mu$ is set to give the correct electron density when the DOS $\rho(E)$ is integrated. $\Sigma(E)$ is the so-called transport distribution function (TDF), which gives access to all the electronic transport properties and is therefore the key quantity to be calculated. $\sigma$ is the thermal average of $\Sigma$ around the Fermi level $\mu$, while $S$ is basically the logarithmic derivative of $\Sigma$ around $\mu$.\cite{Cutler1969} Therefore, high Seebeck coefficients arise from strong, sharp variations in the TDF (i.e. large values of $\vert\frac{d\Sigma}{dE}\vert$). Most theoretical studies of doped thermoelectric materials compute the TDF within the framework of the Boltzmann transport equation with the relaxation time approximation. \cite{Ashcroft} They either consider impurity scattering to be negligible compared to electron-phonon scattering, \cite{Sun2017,Song2017,Parker2010,Peng2011,Yang2015} or estimate the electron-impurity scattering rate by second-order perturbation theory (i.e. Fermi's Golden Rule) using a model description for the impurity scattering.\cite{Wang2011,Zebarjadi2012,Qiu2015} This is reasonable when the doping does not significantly alter the electronic structure and causes only weak electron-impurity scattering, as is the case of La or Nb doped SrTiO$_3$ for instance. \cite{Bouzerar2017} But the whole point of resonant states is that they distort the band structure of the host material and introduce strong scattering. Therefore, in this study, we go beyond the semi-classical Boltzmann formalism to incorporate the full effects of disorder and multiple resonant scattering. We use the Kubo formula expressed in terms of the Green's function $\hat{G}$ of the system \cite{Greenwood1958,Stone1988,Baranger1989,Nikolic2001}
\begin{equation}
\label{kubo}
\Sigma(E) = \frac{\hbar e^2}{\pi \Omega} \; \langle \Tr(\text{Im}\hat{G}(E) \, \hat{v}_x \, \text{Im}\hat{G}(E) \, \hat{v}_x) \rangle,
\end{equation}
brackets denote disorder averaging, $\Omega$ is the total volume, $\hat{v}_x = \frac{it}{\hbar} \sum_{\bf{<i,j>},\sigma} ( x_{\bf{i}} - x_{\bf{j}}) ( c^{\dagger}_{\bf{i},\sigma} c^{}_{\bf{j},\sigma} - c^{\dagger}_{\bf{j},\sigma} c^{}_{\bf{i},\sigma} )$ is the velocity operator along the transport direction $x$, and the Green's function writes
\begin{equation}
\label{Green}
\hat{G}(E) = \frac{1}{E - \hat{H} + i \frac{\gamma_{\text{in}}}{2}}.
\end{equation}
A constant imaginary part has been introduced in the denominator of $\hat{G}(E)$ to account for the presence of inelastic scattering mechanisms in the system, such as electron-phonon (e-ph) or electron-electron (e-e) collisions. It will be further discussed thereafter.  $\Sigma$ can be expressed in terms of the adimensioned TDF $\tilde{\Sigma}$, $\Sigma = \frac{e^2}{a \hbar} \tilde{\Sigma}$, and likewise for the power factor, $\text{PF} = \frac{k_B^2}{a \hbar} \tilde{\text{PF}}$. For definiteness, we set $a = \SI{4}{\angstrom}$, which gives the same volume per atom as in Si or PbTe, leading to $\frac{e^2}{a \hbar}~=~\SI{6.08e3}{S/cm}$ and $\frac{k_B^2}{a \hbar} = \SI{45.18}{\mu W \cdot cm^{-1} \cdot K^{-2}}$.

The exact diagonalization of the Hamiltonian (\ref{Hamiltonian}) would drastically limit the system sizes that could be studied with a reasonable amount of memory and computational time. Therefore, a good alternative to compute $\Sigma(E)$ exactly, fully including vertex corrections, is the Chebyshev-Polynomial Green's Function method (CPGF). \cite{Ferreira2015,Weisse2006} It is a real-space approach particularly suitable for addressing the physics in disordered systems. A brief overview of the method follows. The Green's function $\hat{G}(E)$ is expanded on the Chebyshev polynomials basis:
\begin{equation}
\label{CPGF}
\hat{G}(\tilde{E}) = \sum_{n=0}^\infty g_n \left( \tilde{z} \right) T_n(\tilde{\hat{H}}),
\end{equation}
where $\tilde{z} = \tilde{E} + i \tilde{\gamma}_{\text{in}}/2$ with $\tilde{E}$ and $\tilde{\gamma}_{\text{in}}/2$ rescaled in the energy interval $[-1,1]$, $T_n(\tilde{\hat{H}})$ are the Chebyshev polynomials calculated by the recursion relation $T_{n+1}(\tilde{\hat{H}}) =~2 \tilde{\hat{H}} T_n(\tilde{\hat{H}}) - T_{n-1}(\tilde{\hat{H}})$, and $g_n(z)$ are known complex functions given by
\begin{equation}
g_n(z) = -i \left( 2 - \delta_{n,0} \right) \frac{\left( z - i \sqrt{1 - z^2} \right)^n}{\sqrt{1 - z^2}}.
\end{equation}
Inserting equation (\ref{CPGF}) into equation (\ref{kubo}) yields
\begin{equation}
\Sigma(\tilde{E}) = \frac{\hbar e^2}{\pi \Omega} \sum_{n,n'} \mu_{n,n'} \, \text{Im} g_n(\tilde{z}) \, \text{Im} g_{n'}(\tilde{z}),
\end{equation}
with 
\begin{equation}
\label{moments}
\mu_{n,n'} = \langle \Tr(T_n(\tilde{\hat{H}}) \, \tilde{\hat{v}}_x \, T_{n'}(\tilde{\hat{H}}) \, \tilde{\hat{v}}_x) \rangle.
\end{equation}
The quantities $\mu_{n,n'}$ are called the moments and are calculated by iterative multiplications of the Hamiltonian matrix operator. Due to the superior convergence properties of the Chebyshev polynomials \cite{boyd01} and to the presence of the inelastic self-energy, only a finite number of terms in equation~(\ref{CPGF}) need to be computed in order to obtain the exact Green's function of the system. Around $1800 \times 1800$ moments are sufficient for the TDF to fully converge. Periodic boundary conditions are used to reach the thermodynamic limit more easily. Here, we compute the TDF on systems of size $N = 1200 \times 200 \times 200$ (\num{48e6} sites), this slab geometry allows faster convergence. The trace in equation~(\ref{moments}) is evaluated efficiently by a stochastic method involving random vectors as described in Ref.\onlinecite{Weisse2006} and Ref.\onlinecite{Ferreira2015}. When calculating the TDF for such a large system size, only a few random vectors and disorder configurations are necessary. We have checked that the clean limit is perfectly recovered for both open and periodic boundary conditions.

Regarding $\gamma_{\text{in}}$ in equation~(\ref{Green}), it can be interpreted as the inelastic contribution to the electron relaxation time. \cite{Mahan} In thermoelectric materials, scattering rates at room temperature typically range from $\SI{1}{meV}$ to $\SI{100}{meV}$. In certain Half-Heuslers such as ZrNiSn, for instance, particularly weak e-ph couplings lead to inelastic scattering rates varying between $\SI{1}{meV}$ and $\SI{20}{meV}$ \cite{Zhou2018} while strong e-e scattering in SrTiO$_3$ leads to $\gamma_{\text{in}}$ ranging from $\SI{50}{meV}$ to $\SI{200}{meV}$ and even higher. \cite{Bouzerar2017} In between, scattering rates vary from $\SI{10}{meV}$ to $\SI{100}{meV}$ in Si \cite{Witkoske2017,Zhang2018,Qiu2015} or from $\SI{20}{meV}$ to $\SI{60}{meV}$ in pristine and doped SnSe. \cite{MELENDEZ201870} From the bandwidth in these compounds, we estimate $t$ ranging from $\SI{0.3}{eV}$ to $\SI{1}{eV}$, therefore we will consider  $0.02 \, t \le \gamma_{\text{in}} \le 0.2 \, t$. This corresponds to $\gamma_{\text{in}}$ ranging from $\SI{10}{meV}$ to $\SI{200}{meV}$ and mean free paths between $\SI{50}{\angstrom}$ and $\SI{500}{\angstrom}$ in the pristine case, which is well in line with calculated values in PbTe, \cite{Song2017} for instance. As we will see shortly, a small inelastic scattering is most favorable for resonant enhancements of the PF, therefore we set $\gamma_{\text{in}} = 0.02 \, t$ unless specified otherwise. The choice of a constant $\gamma_{\text{in}}$ preserves the generality of this investigation, since incorporating energy and temperature dependences requires a material-specific study. Note also that computations for smaller $\gamma_{\text{in}}$ would imply a strong increase in the number of calculated moments before convergence is reached.


\section{Numerical results and discussion}

\begin{figure}
\vspace{-5pt}
\includegraphics[width=0.95\columnwidth]{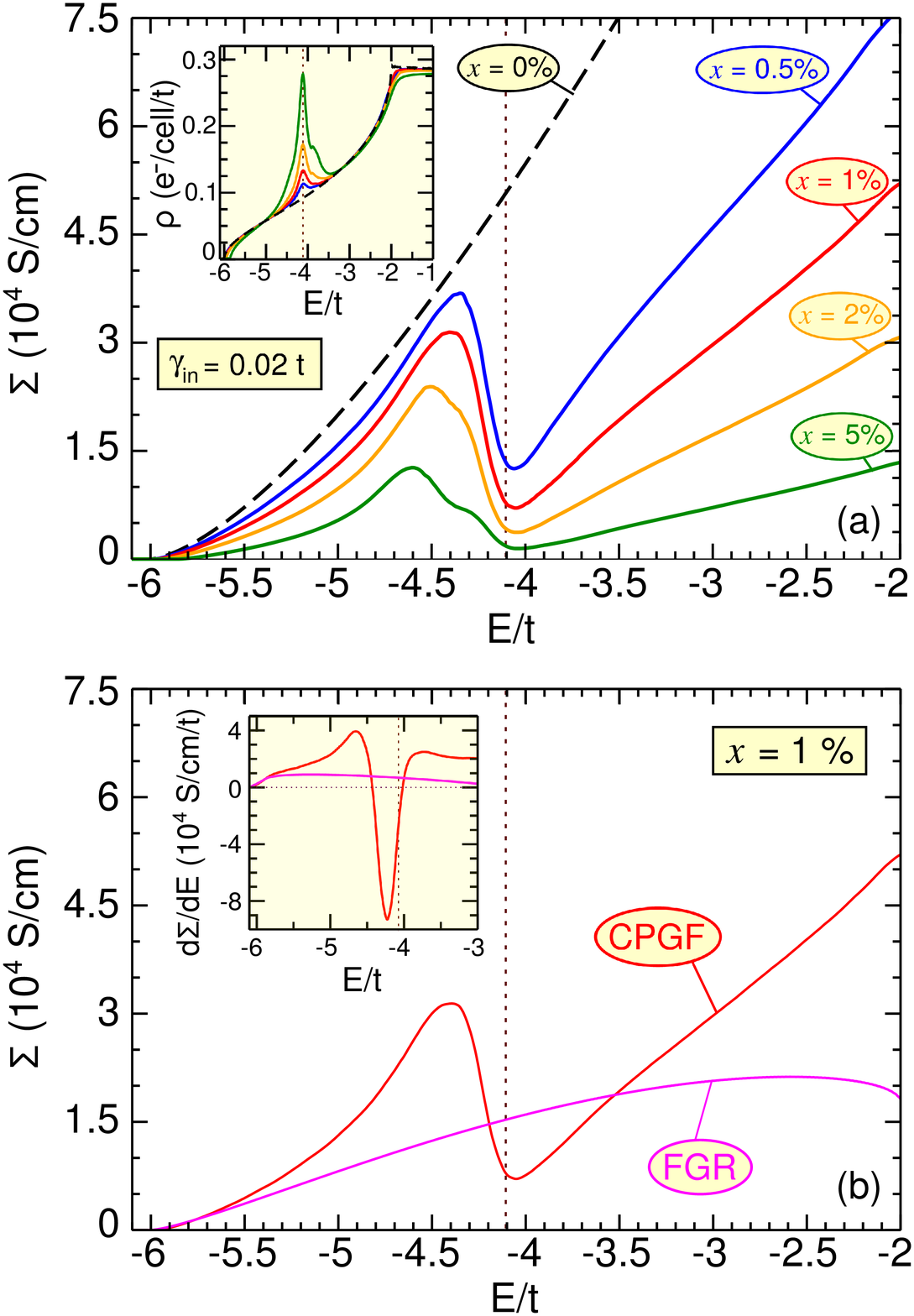}
\caption{(a) Transport distribution function $\Sigma(E)$ and density of states $\rho(E)$ (inset) for five impurity concentrations, from $x=0\%$ (reference, black dashed line) to $x=5\%$. A vertical dashed line marks the position of the resonant peak. (b) $\Sigma(E)$ and its derivative (inset) for $x= 1\%$ calculated exactly (CPGF) and by second-order perturbation theory (FGR).}
\label{fig2}
\end{figure}

In what follows, we consider the set of values $\epsilon = -4 \, t$ and $V = 0.3 \, t$ that were found optimal in a previous study that completely ignored disorder. \cite{Thebaud2017} Unless specified otherwise, the electronic properties will be calculated using these values. Fig.~\ref{fig2}(a) shows the TDF and the DOS (inset) for different impurity concentrations $x = \frac{N_\text{imp}}{N}$, with $N_\text{imp}$ the number of randomly distributed defects. Five concentrations are considered, from $x=0\%$ (the pristine reference case) to $x=5\%$. The defects introduce a local peak in the DOS, which is considered the main signature of resonant states in the literature, and the mechanism by which the transport properties are enhanced. The higher the impurity concentration, the bigger and sharper the peak, so we would expect the best thermoelectric performances from the highest concentrations. As we will see shortly, this is not the case. The resonant peak gives rise in the TDF to a sharp, asymmetrical dip, as the extended states acquire a more localized character by hybridizing with impurity states. At high defect concentrations, electron transport is more suppressed across the whole energy range so the variations of the TDF are gentler ($\vert\frac{d\Sigma}{dE}\vert$ is reduced as $x$ increases). We compare in Fig.~\ref{fig2}(b) the TDF and $\frac{d\Sigma}{dE}$ calculated by CPGF and by the often used Fermi's Golden Rule (FGR). Matthiessen's rule states that the total scattering rate is $\gamma_\text{tot} = \gamma_\text{imp} + \gamma_\text{in}$ with $\gamma_\text{imp}$ the impurity scattering rate. FGR leads to $\gamma_\text{imp}(E) = 2 \pi x \epsilon^2 \rho_0(E)$, $\rho_0(E)$ being the DOS of the clean system. The FGR transport distribution function is given by $\Sigma_{\text{\tiny FGR}}(E) = \frac{\gamma_\text{in}}{\gamma_\text{tot}} \Sigma^{(0)}(E)$ where $\Sigma^{(0)}(E)$ is the pristine TDF. Notice that the exact $\Sigma(E)$ cannot be cast into such an analytical form. Clearly, the FGR approach completely fails to give the correct dependance of the TDF. In particular, the dip is entirely absent. The discrepancy is even worse for the derivatives, which are directly linked to the Seebeck coefficients. Therefore, when resonant states are involved, second-order perturbation theory breaks down.

\begin{figure}
\includegraphics[width=0.93\columnwidth]{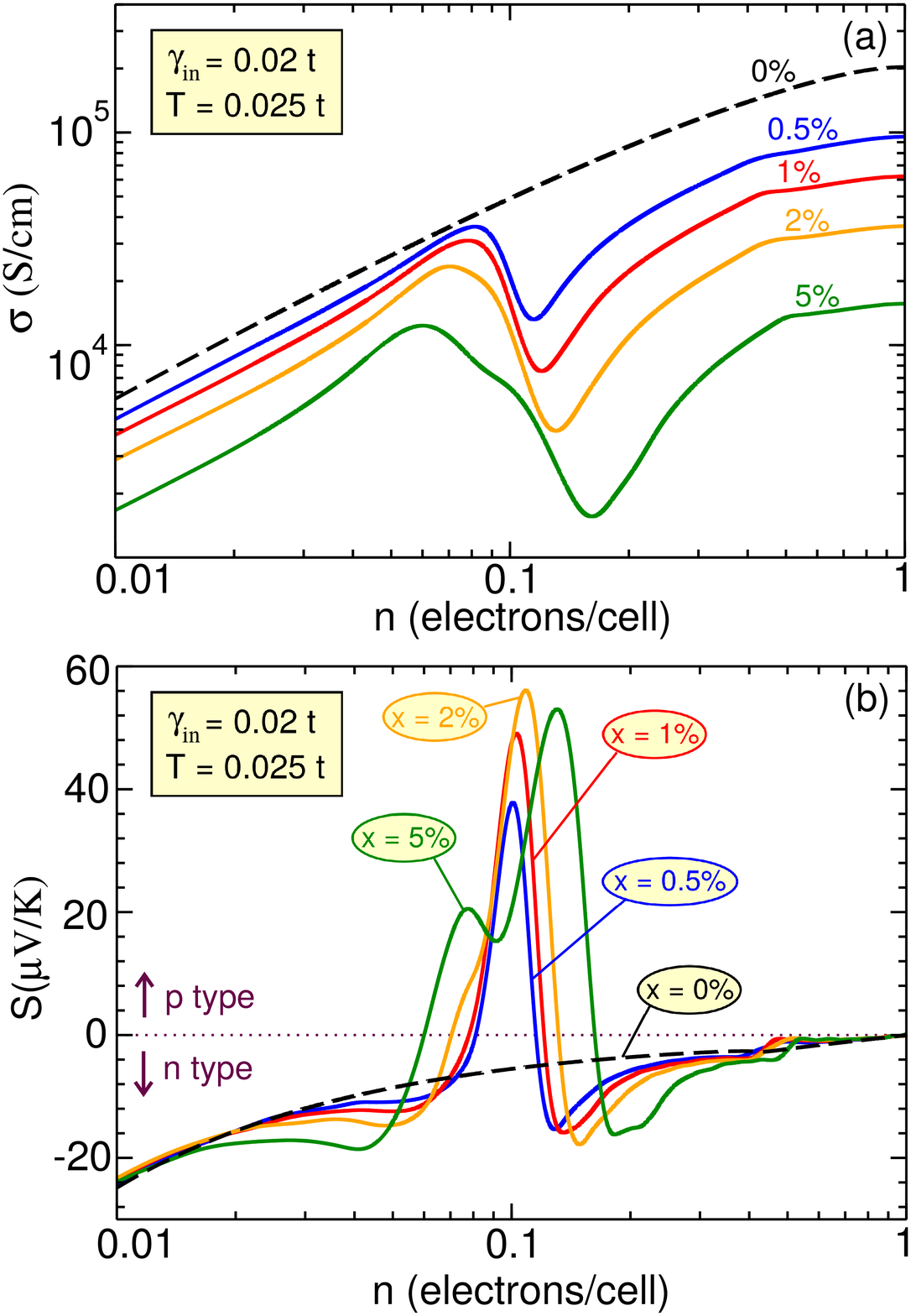}
\caption{(a) Electrical conductivity $\sigma$ and (b) Seebeck coefficient $S$ as a function of the electron density $n$ for five impurity concentrations, from $x=0\%$ to $x=5\%$.}
\label{fig3}
\end{figure}

From the results of Fig.~\ref{fig2}(a) and equation (\ref{conductivite}) and (\ref{Seebeck}), we compute the room-temperature electrical conductivity $\sigma$ and Seebeck coefficient $S$, as plotted in Fig.~\ref{fig3} as a function of the electron density $n$. $T$ is set to $0.025 \, t$ which corresponds to room temperature if $t \approx \SI{1}{eV}$. $\sigma$ is reduced by the disorder and still exhibits the same features as $\Sigma(E)$ ($T$ being relatively small). This reduction would be detrimental to the PF, but the sharp variations in the TDF lead to a boost of the Seebeck coefficient that overcompensates the suppression of $\sigma$. This is accompanied by a sign inversion of $S$ around $n = \SI{0.1}{electrons/cell}$. Thus, $S$ can change sign in the disordered systems, while it remains n-type in the absence of resonant states. This interesting feature opens the possibility of changing the thermoelectric material from n-type to p-type just by introducing the appropriate impurity or dopant. Therefore, one could build a device with both n and p legs from the same semiconductor host. This would be advantageous for device performance and reliability, provided that the PF is sufficiently large. 

\begin{figure}
\includegraphics[width=0.95\columnwidth]{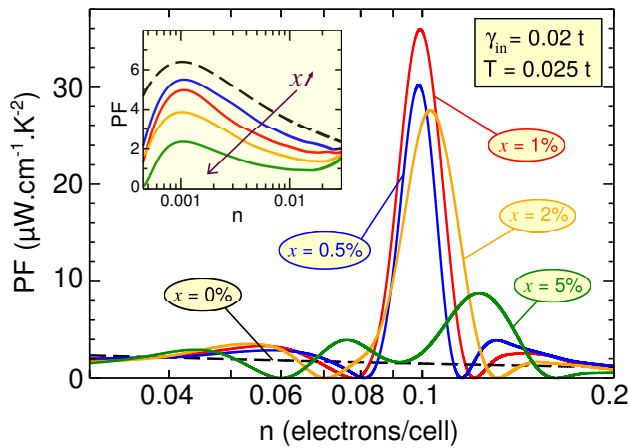}
\caption{Power factor $\sigma S^2$ as a function of the electron density $n$ for five impurity concentrations, from $x=0\%$ to $x=5\%$. Inset: PF for lower electron densities.}
\label{fig4}
\end{figure}

The PF is plotted in Fig.~\ref{fig4} as a function of $n$. The pristine system exhibits a maximum of $\SI{6.3}{\mu W \cdot cm^{-1} \cdot K^{-2}}$ around $n = \SI{e-3}{electrons/cell}$, corresponding to a conductivity of $\SI{400}{S/cm}$ and a Seebeck coefficient of $\SI{-130}{\mu V/K}$. Note that this relatively low value of the power factor is partly due to the absence of band degeneracy and anisotropy in our single-band model. The effects of resonant impurities relative to the pristine case would not be affected by band degeneracy, while the case of anisotropic orbitals will be examined thereafter. When resonant defects are introduced, the PF is suppressed at low densities (inset), because multiple impurity scatterings have a stronger effect on the long wavelength carriers. By contrast, around $n = \SI{0.1}{electrons/cell}$, the PF now exhibits a large enhancement due to the boost of the Seebeck coefficient that overcompensates the drop in conductivity. The largest increase corresponds to $x = 1 \%$, for which the PF reaches its maximum $\SI{35.9}{\mu W \cdot cm^{-1} \cdot K^{-2}}$, a sixfold enhancement compared to that of the clean system. For $x = 5 \%$, the boost is still present but less spectacular (a ratio less than 2) due to the gentler variations in the TDF. This is an important and surprising finding: to achieve an efficient enhancement of the thermoelectric properties with resonant states, the defect concentration should be kept relatively low, typically around $1 \%$. From an experimental point of view, that is favorable, because such concentrations usually lie below the solubility limit. \cite{Boeisenko1987} Co-doping with a donor atom acting as an electron reservoir is necessary to shift the Fermi level inside the resonant peak, where the PF is enhanced and the Seebeck inversion occurs. This carrier density optimization still requires at most $10 \%$ co-doping, which is reasonable. We define $\text{PF}_{\text{max}}$ as the optimum PF with respect to the carrier concentration. $\text{PF}_{\text{max}}$ extracted from Fig.~\ref{fig4} are shown in the first row of table~\ref{PFmax}.

\begin{table}
\vspace{5pt}
\setlength{\tabcolsep}{6pt}
\resizebox{\columnwidth}{!}{
\begin{tabular}{ m{40pt} c c c c c} 
\toprule
 & $x=0\%$ & $0.5\%$ & $1\%$ & $2\%$ & $5\%$ \\
\midrule
$\epsilon = -4 \, t$ \newline $V= 0.3 \, t$ & 6.3 & 30.2 & 35.9 & 27.6 & 8.7 \\
\midrule
$\epsilon = -5.5 \, t$ \newline $V= 0.3 \, t$ & 6.3 & 1.3 & 1.7 & 1.8 & 1.8 \\
\midrule
$\epsilon = -4 \, t$ \newline $V=\, t$ & 6.3 & 0.01 & 0.05 & 0.06 & 0.05 \\
\bottomrule
\end{tabular}
}
\caption{Room-temperature optimum power factor in \si{\mu W \cdot cm^{-1} \cdot K^{-2}} for $\gamma_{\text{in}}~=~0.02 \, t$ and several values of the on-site potential and hybridization parameter.}
\label{PFmax}
\end{table}

\begin{figure}
\vspace{-4pt}
\includegraphics[width=1.0\columnwidth]{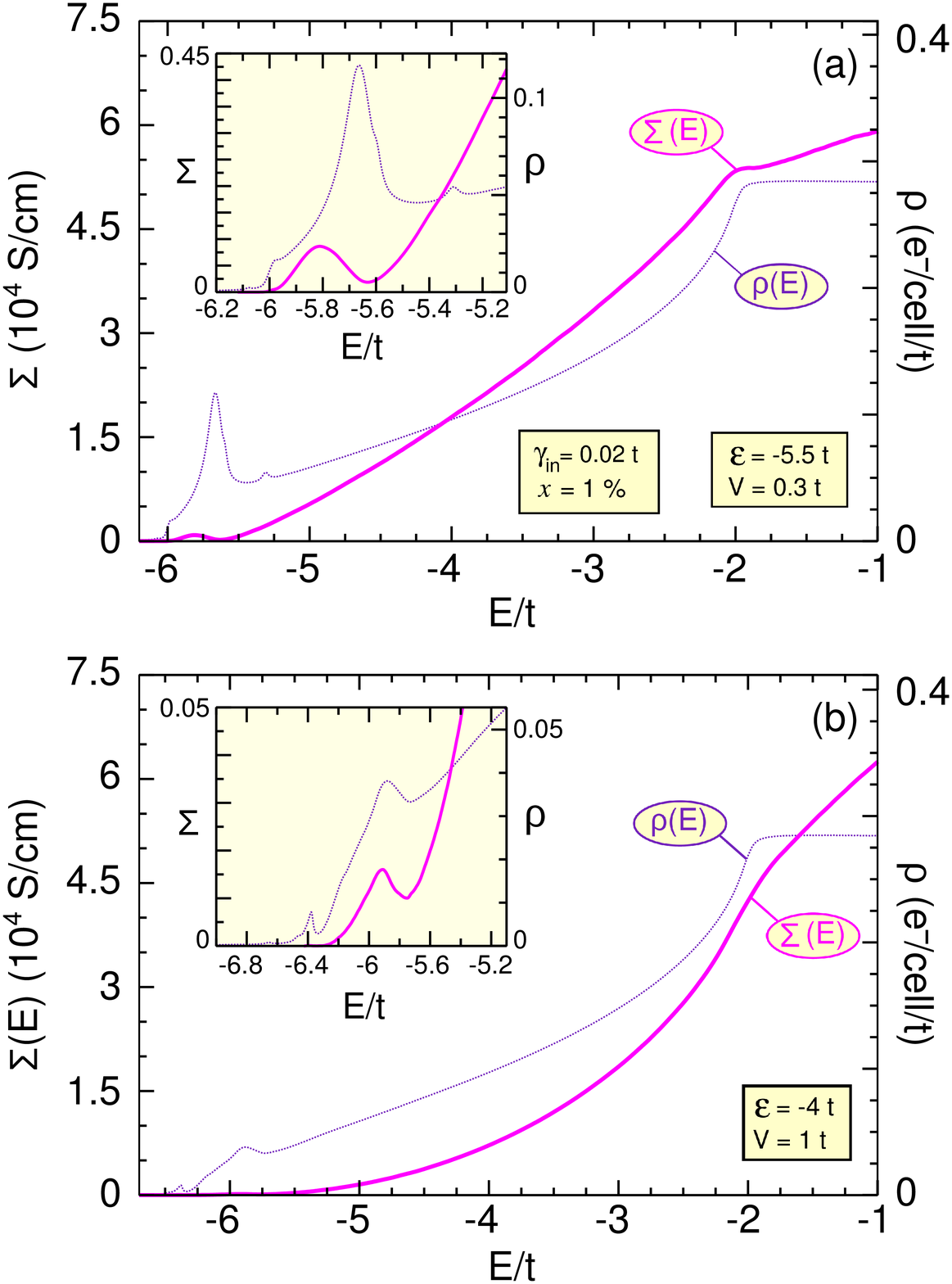}
\caption{Density of states $\rho(E)$ (dotted line, right axis) and transport distribution function $\Sigma(E)$ (left axis) for $x = 1\%$ with (a) $\epsilon = -5.5 \, t$, $V= 0.3 \, t$ and (b) $\epsilon = -4 \, t$, $V= \, t$. Inset: zoom on the bottom of the conduction band, notice the different scales for the TDF.}
\label{fig5}
\end{figure}

We now address the influence of both $\epsilon$ and $V$ on the transport properties. In Fig.~\ref{fig5}(a), the DOS and TDF are plotted for $x = 1\%$ resonant impurities with $\epsilon = -5.5 \, t$ and $V= 0.3 \, t$. The increase of the on-site potential shifts the position of the resonant peak much closer to the bottom of the conduction band. The resulting TDF also features a much smaller dip in $\Sigma(E)$ at the position of the peak in $\rho(E)$. Consequently, $\vert\frac{d\Sigma}{dE}\vert$ remains quite weak, and so does the Seebeck coefficient. There is still a sign inversion, but no boost in the PF. Indeed it is even suppressed by a factor 3-4 with respect to the reference value of $\SI{6.3}{\mu W \cdot cm^{-1} \cdot K^{-2}}$ (see the second row of table~\ref{PFmax}). Thus the resonant peak should not be too close to the band edge, but deep inside the conduction band. We now focus on the effect of a larger hybridization, which implies a stronger coupling between conduction and defect states. Results are depicted in Fig.~\ref{fig5}(b). The increase in hybridization also pushes the resonant states at the very edge of the band, and severely suppresses the TDF below $-5 \, t$ (notice the scale in the inset). There is still a small dip in $\Sigma(E)$ and a sign inversion of $S$, but because the carriers are now so localized in this energy range, the PF shrinks by at least two orders of magnitude compared to that of the reference. For $x= 1 \%$, $\text{PF}_{\text{max}}$ is now $\SI{0.05}{\mu W \cdot cm^{-1} \cdot K^{-2}}$ (see the third row of table~\ref{PFmax} for the other concentrations). This suppression of the PF is entirely due to a huge reduction in the conductivity caused by multiple scattering events that become important at low energy in the presence of stronger disorder. These findings are consistent with the results obtained in Ref.\onlinecite{PhysRevB.97.205203} for Tl-doped PbTe, in which the Tl doping creates a resonant bump at the edge of the valence band associated with a much higher resistivity compared to that of Na doping which behaves as a reservoir. It should also be mentioned that resonant states formed by antisites in Fe$_2$VAl have been found to suppress the PF by more than an order of magnitude while changing the sign of $S$.\cite{Bilc2011} The takeaway to obtain a boost of the PF is that the substituting element should be suitably selected in order to create a resonant peak far from the band edge. This could also explain why many claims of experimental enhancement of the PF by resonant states remain controversial, and why no sign inversion of the Seebeck coefficient has been observed so far. These effects are indeed sensitive to the hybridization, on-site potential and position of the Fermi level. Additionally, it is difficult to rule out other enhancement mechanisms, such as energy filtering effects resulting from ionized impurity scattering, for instance.

\begin{figure}
\includegraphics[width=0.93\columnwidth]{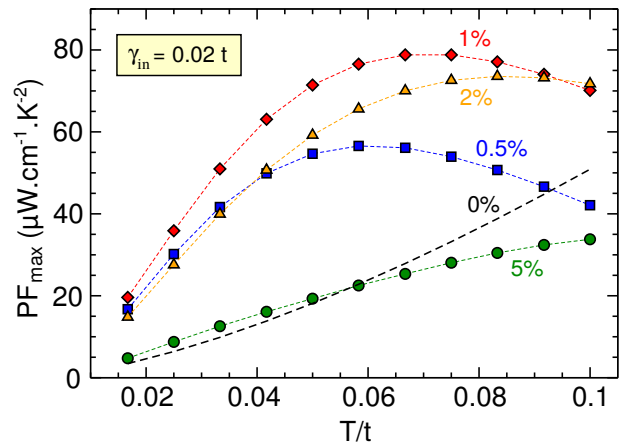}
\caption{Optimum power factor as a function of temperature for five impurity concentrations, from $x=0\%$ to $x=5\%$.}
\label{fig6}
\end{figure}

Since thermoelectric materials are meant to be used in a wide range of temperature, we now discuss the $T$-dependence of $\text{PF}_{\text{max}}$. Fig.~\ref{fig6} shows $\text{PF}_{\text{max}}$ as a function of temperature for the same impurity concentrations as in Fig.~\ref{fig4}. It increases when the temperature rises, reaching a broad maximum, and then decreases slowly in the disordered systems. This high temperature behaviour results from the sharp variations in the TDF being smoothed out by the thermal average. Accordingly, at high values of $x$, the variations in the TDF are broader and are less sensitive to the thermal average, so the maximum region is shifted to higher temperatures. An important finding is that $\text{PF}_{\text{max}}$ itself is robust, suggesting that resonant states could be efficient for both low and high temperature power generation.

\begin{figure}
\includegraphics[width=1.0\columnwidth]{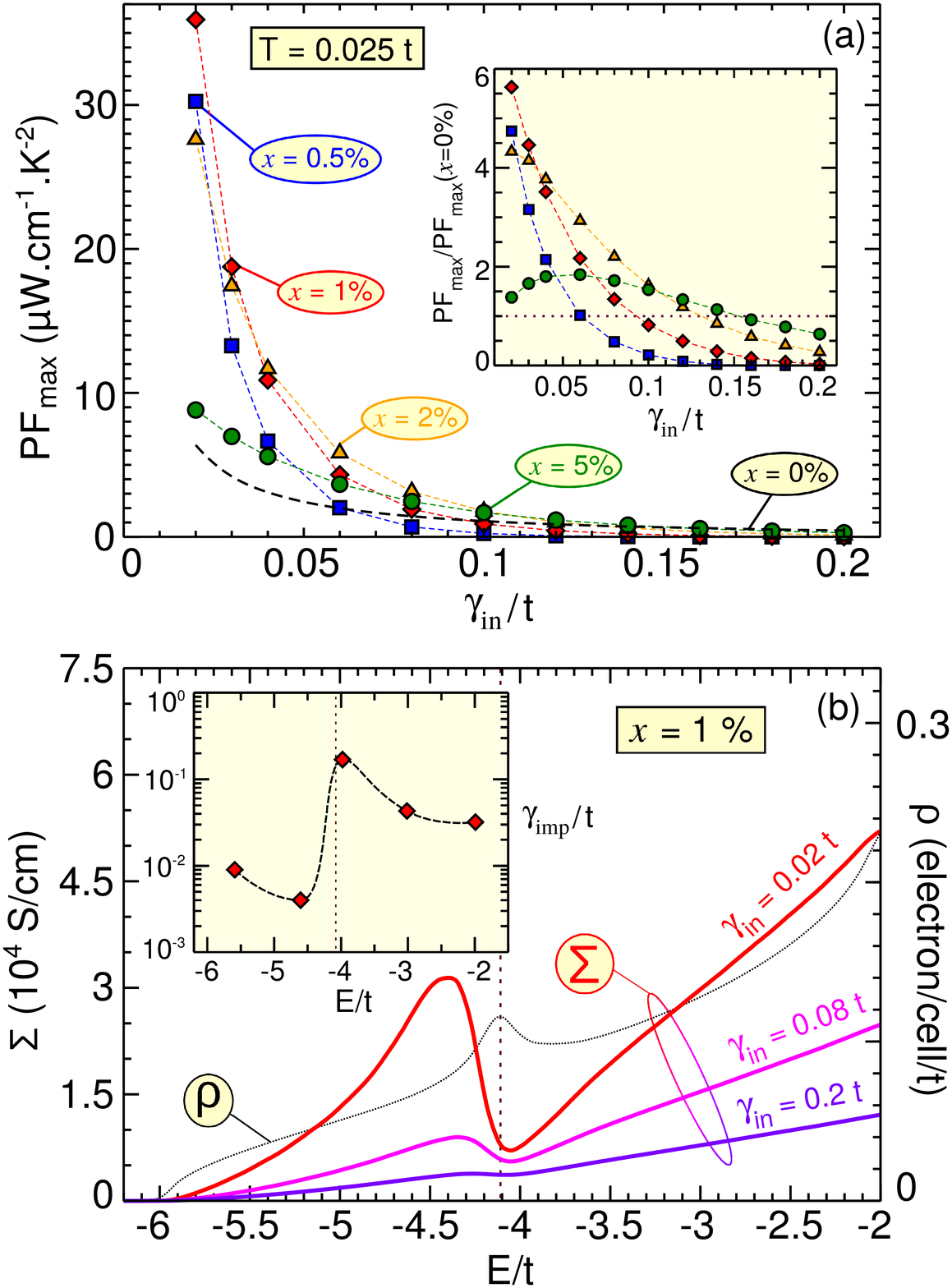}
\caption{(a) Optimum power factor as a function of the inelastic scattering rate $\gamma_{\text{in}}$ for five impurity concentrations, from $x=0\%$ to $x=5\%$. Inset: Ratio of the optimum power factor with respect to that of the clean system. (b) $\Sigma(E)$ for $x = 1\%$ with $\gamma_\text{in} = 0.02 \, t$, $0.08 \, t$ and $0.2 \, t$ (left axis). $\rho(E)$ is also shown (dotted line, right axis). A vertical dashed line marks the position of the resonant peak. Inset: the calculated impurity scattering rate along the $\Gamma$-X direction as a function of energy. The dashed curve is a guide to the eye.}
\label{fig7}
\end{figure}

We now propose to investigate the role of inelastic scattering. In Fig.~\ref{fig7}(a), we plot $\text{PF}_{\text{max}}$ as a function of $\gamma_{\text{in}}$. Clearly, the PF is completely suppressed if the inelastic scattering is too strong. This results from the competition between resonant impurity scatterings (elastic processes) and inelastic scatterings. In Fig.~\ref{fig7}(b) is presented the TDF for $\gamma_\text{in} = 0.02 \, t$, $ 0.08 \, t$ and $0.2 \, t$ with $x = 1 \%$. The inset shows the impurity scattering rate $\gamma_\text{imp}$, extracted from an analysis of the single-particle spectral function $A({\bf q},E) = -\frac{1}{\pi} \langle \text{Im} G({\bf q},E) \rangle$.
$G({\bf q},E)$ is the spatial Fourier transform of the disordered Green's function $G_{\bf{ij},\sigma}(E) = \langle {\bf{i}} \sigma| \hat{G}(E) | {\bf{j}} \sigma\rangle$, where ${|\bf{i}\sigma\rangle}$ is the real-space basis ($\bf{i}$ runs over the lattice sites). $\gamma_\text{imp}$ exhibits a non-monotonic behavior and large variations across the resonant peak, from $\num{4e-3} \, t$ at $E = -4.5 \, t$ to $\num{2e-1} \, t$ at $-4 \, t$. At the position of the dip in the TDF, where the electronic states have a stronger localized character, transport is not very sensitive to the strength of $\gamma_\text{in}$ because $\gamma_\text{imp} \approx \num{2e-1} \, t$ dominates. In contrast, if $\gamma_\text{imp}$ is smaller than $\gamma_\text{in}$, which is the case for states associated with large values of the TDF ($\gamma_\text{imp} \approx \num{4e-3} \, t$ at $E = -4.5 \, t$), then increasing $\gamma_\text{in}$ strongly suppresses $\Sigma(E)$. Thus, large inelastic scattering rates have the overall effect of reducing the disparities in the TDF, leading to poor values of the Seebeck coefficient. If we now consider small values of $\gamma_{\text{in}}$ (below $0.03 \, t$) for $x \le 2\%$, we observe a huge increase of $\text{PF}_{\text{max}}$ as $\gamma_\text{in}$ is reduced. If we extrapolate to $\gamma_\text{in} \le 0.02 \, t$ for $x = 1\%$, an enhancement factor of more than an order of magnitude could even be reached. Hence, due to the competition between elastic and inelastic scattering, the impurity concentration should be tuned with respect to the inelastic scattering rate in the host material to reach an optimal boost of the PF. Compounds exhibiting strong e-ph or e-e scattering should not be the best candidates for resonant substitution doping. 

\begin{figure}
\vspace{10pt}
\includegraphics[width=1.0\columnwidth]{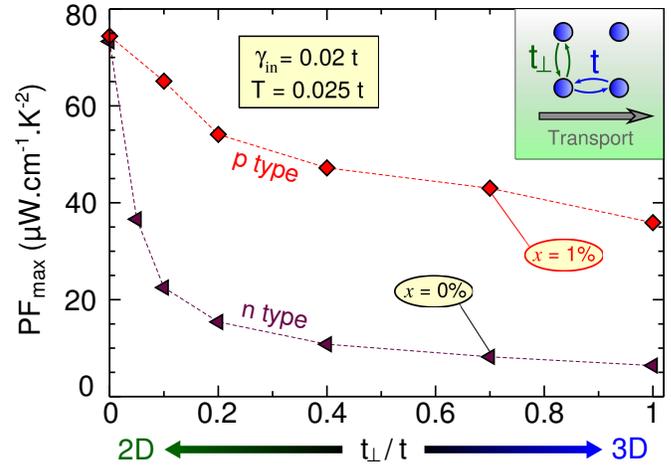}
\caption{Optimum power factor as a function of the anisotropy ratio $t_\perp / t$ for $x=0\%$ (reference) and $x=1\%$.}
\label{fig8}
\end{figure}

Till now, we have been considering an isotropic electronic structure (s-type orbitals), but it is worth considering the influence of orbital anisotropy.\cite{Parker2013,Bilc2015} Low-dimensional confinement is expected to introduce sharp structures in the DOS and thus sharp variations in the TDF, thereby boosting the Seebeck coefficient. To evaluate the gain in the PF that could be obtained from resonant states in anistropic systems, we now introduce a different hopping $t_\perp$ in a direction perpendicular to transport. The optimum PF for the reference $x = 0 \%$ and for $x=1\%$ is presented in Fig.~\ref{fig8} as a function of $t_\perp / t$. First notice that the PF of the pristine system strongly increases with the anisotropy, from $\SI{6.3}{\mu W \cdot cm^{-1} \cdot K^{-2}}$ (3D) to $\SI{73.3}{\mu W \cdot cm^{-1} \cdot K^{-2}}$ (2D). This confirms that two-dimensional confinement in itself does favor good performances. The maximum PF (p-type) for $x=1\%$ also increases with the anisotropy, from $\SI{35.9}{\mu W \cdot cm^{-1} \cdot K^{-2}}$ (3D) to $\SI{74.4}{\mu W \cdot cm^{-1} \cdot K^{-2}}$ (2D). Surprisingly, for $t_\perp = 0$ we find no boost in the PF, suggesting that the presence of resonant states in fully confined systems might not further enhance the thermoelectric properties. However, one should emphasize that even for finite but low ratios $t_\perp/t$ (down to 0.05), the PF can be significantly increased by resonant states. This is promising for bulk systems in which charge carriers populate highly anisotropic orbitals. This is, for instance, the case of n-doped SrTiO$_3$, in which the Titanium 3d orbitals exhibit a $t_\perp/t \approx 0.1$. \cite{Zhong2012,Bouzerar2017} Moreover, using resonant states in fully confined materials could be interesting for the sign inversion of $S$ alone. 

To conclude, we have used the Chebyshev-Polynomial Green's Function method to address the effects of resonant impurities on electron transport. Although resonant states suppress the electrical conductivity, they may also lead to a boost and a sign inversion of the Seebeck coefficient. Consequently, the power factor can increase by one order of magnitude. However, the resonant peak should be located far from the band edge, otherwise the thermoelectric performances are destroyed. Additionally, the optimal boost of the power factor depends crucially on the interplay between elastic and inelastic scattering. Strong electron-phonon or electron-electron scattering are found to preclude the possibility of enhancing thermoelectric transport. Therefore, materials featuring long electron mean free paths and weak inelastic scattering, such as PbTe, \cite{Song2017} certain Half-Heuslers compounds \cite{Zhou2018} or even graphene \cite{Morozov2008,CastroNeto2009} should be promising candidates. Finally, the resonant boost of the power factor is found robust in the case of anisotropic orbitals. This study will hopefully contribute to a better understanding of resonant states in the context of thermoelectric power generation. Our methodology is very general and can be combined with realistic first-principles calculations to guide experimentalists and pave the way in the search for compounds exhibiting resonant effects.

\begin{footnotesize}
\bibliographystyle{plain} 

\end{footnotesize}

\end{document}